\documentclass[prl,unsortedaddress,twocolumn,superscriptaddress,10pt]{revtex4}

\usepackage[utf8]{inputenc}
\usepackage[english]{babel}
\usepackage{amsmath}
\usepackage{amsfonts}
\usepackage{amssymb}
\usepackage{graphicx}
\usepackage{bm}
\usepackage{spjheadersSinglePageNumTotal}

\usepackage{graphicx}
\DeclareGraphicsExtensions{ps,eps,ps.gz,frh.eps,ml.eps}
\usepackage{epstopdf}
\begin{document}


\newcommand{\Atilde}{\widetilde{A}}
\newcommand{\Etilde}{\widetilde{E}}
\newcommand{\w}{\omega}
\newcommand{\wopt}{\omega_{\rm opt}}
\newcommand{\wthz}{\omega_{\rm THz}}
\newcommand{\W}{\Omega}
\newcommand{\boldMatrix}[1]{ \underline{\underline{\mathbf{#1}}} }
\newcommand{\boldVect}[1]{ \underline{\mathbf{#1}} }
\newcommand{\chitwo}{\chi^{(2)}}
\newcommand{\chithree}{\chi^{(3)}}
\newcommand{\LN}{LiNbO$_3$} 

\received{}
\accepted{}

\Header{Sub-luminal THz for particle acceleration}{}


\title{Demonstration  of sub-luminal propagation of single-cycle terahertz pulses for particle acceleration}

\author{D. A. Walsh} 
\affiliation{Accelerator Science and Technology Centre, Science and Technology Facilities Council, Daresbury Laboratory, Keckwick Lane, Daresbury, Warrington WA4 4AD, United Kingdom}
\affiliation{The Cockcroft Institute, Sci-Tech Daresbury, Keckwick Lane, Daresbury, Warrington WA4 4AD, United Kingdom}
\author{D. S. Lake}

\affiliation{The Cockcroft Institute, Sci-Tech Daresbury, Keckwick Lane, Daresbury, Warrington WA4 4AD, United Kingdom}
\affiliation{School of Physics and Astronomy \& Photon Science Institute, The University of Manchester, Manchester M13 9PL, United Kingdom}
\author{E. W. Snedden} 
\affiliation{Accelerator Science and Technology Centre, Science and Technology Facilities Council, Daresbury Laboratory, Keckwick Lane, Daresbury, Warrington WA4 4AD, United Kingdom}
\affiliation{The Cockcroft Institute, Sci-Tech Daresbury, Keckwick Lane, Daresbury, Warrington WA4 4AD, United Kingdom}
\author{M. J. Cliffe}
\author{D. M. Graham}
\affiliation{The Cockcroft Institute, Sci-Tech Daresbury, Keckwick Lane, Daresbury, Warrington WA4 4AD, United Kingdom}
\affiliation{School of Physics and Astronomy \& Photon Science Institute, The University of Manchester, Manchester M13 9PL, United Kingdom}
\author{S. P. Jamison}
	\email[]{steven.jamison@stfc.ac.uk}
\affiliation{Accelerator Science and Technology Centre, Science and Technology Facilities Council, Daresbury Laboratory, Keckwick Lane, Daresbury, Warrington WA4 4AD, United Kingdom}
\affiliation{The Cockcroft Institute, Sci-Tech Daresbury, Keckwick Lane, Daresbury, Warrington WA4 4AD, United Kingdom}


\date{\today}

\begin{abstract}
The sub-luminal phase velocity of electromagnetic waves in free space is generally unobtainable, being closely linked to forbidden faster than light group velocities. The requirement of effective sub-luminal phase-velocity in laser-driven particle acceleration schemes imposes a fundamental limit on the total acceleration achievable in free-space, and necessitates the use of dielectric structures and waveguides for extending the field-particle interaction. Here we demonstrate a new  travelling-source and free space propagation approach to overcoming the sub-luminal propagation limits. The approach exploits the relative ease of generating ultrafast optical sources with slow group velocity propagation, and a group-to-phase front conversion through non-linear optical interaction near a material-vacuum boundary. The concept is demonstrated with two terahertz generation processes, non-linear optical rectification and current-surge rectification. The phase velocity is tunable, both above and below vacuum speed of light $c$, and we report measurements of longitudinally polarized electric fields propagating between $0.77c$ and $1.75c$. The ability to scale to multi-MV/m field strengths is demonstrated.
Our approach paves the way towards the realization of cheap and compact particle accelerators with unprecedented femtosecond scale control of particles.
\end{abstract} 

\maketitle

Femtosecond duration relativistic electron beams are in demand for ultrafast electron diffraction, as sources of femtosecond x-rays in free-electron laser facilities such as LCLS\,\cite{EmmaLCLS2010} and SACLA\,\cite{Shintake2012}, and for future lepton colliders such as CERNs proposed 35\,km long 
Compact Linear Collider, CLIC\,\cite{CLICCDR}.
For over 50 years GHz frequency electromagnetic fields have been the drivers in increasingly sophisticated generations of particle accelerators, but face significant  challenges in meeting the demands for control and manipulation of particle beams on the femtosecond scale. Physical limits in the achievable acceleration field strengths at GHz frequencies also underlies the km-scale and multi-MW power consumption of high energy accelerators.
Terahertz (THz) driven acceleration is poised to revolutionise particle accelerators for femtosecond duration relativistic beams. 
With their picosecond period fields, THz sources offer the ability to capture and finely control few-fs particle beam~\cite{Huang2015,Jamison2012,Hebling2011}, while the few-ps envelope of energy content offers orders of magnitude improvement in matching energy localization to the charged particles being accelerated.
A THz acceleration scheme has recently demonstrated the ability to accelerate non-relativistic particles, using a waveguide structure to accelerate electrons from an initial energy of 60\,keV to 67\,keV over a distance of 3\,mm \cite{Nanni2015}. Ultrafast THz pulse sources have also recently been shown to produce GV/m electric field strengths \cite{Shalaby2014,Vicario2014,VicarioPRL2014}, exceeding by an order of magnitude that of normal-conducting radio-frequency (RF) accelerators \cite{Wang2011}, and two orders of magnitude that of superconducting accelerators such as the European XFEL \cite{Kostin2009}. To obtain continuous and length-scalable acceleration of relativistic particles however, requires phase matching of the electromagnetic carrier wave with the $\beta<1$ velocity of the particles in order that these high electric field strengths can be exploited. 

The highest field strength THz sources are obtained in near-single-cycle electromagnetic pulses, containing an exceptionally broadband coherent spectrum. The single-cycle nature of the pulse is both central to its application for particle acceleration, and to-date the principle obstacle to scalable implementation. The single-cycle near transform limited time profile gives rise to the highest field strengths, and the 100\,fs to 1\,ps carrier period is ideally matched to capture and control particle beams with sub 10\,fs duration; additionally the broadband coherence allows for customized electric field temporal profiles, such as linear ramps. With an electromagnetic energy temporally concentrated within 1\,ps and well matched to the duration of particle bunches, orders of magnitude efficiency gains over conventional RF acceleration (with the $\mu$s and ms fills times of normal and superconducting cavities, respectively) may be achieved.

\begin{figure*}[htb!]
\includegraphics[width=17cm]{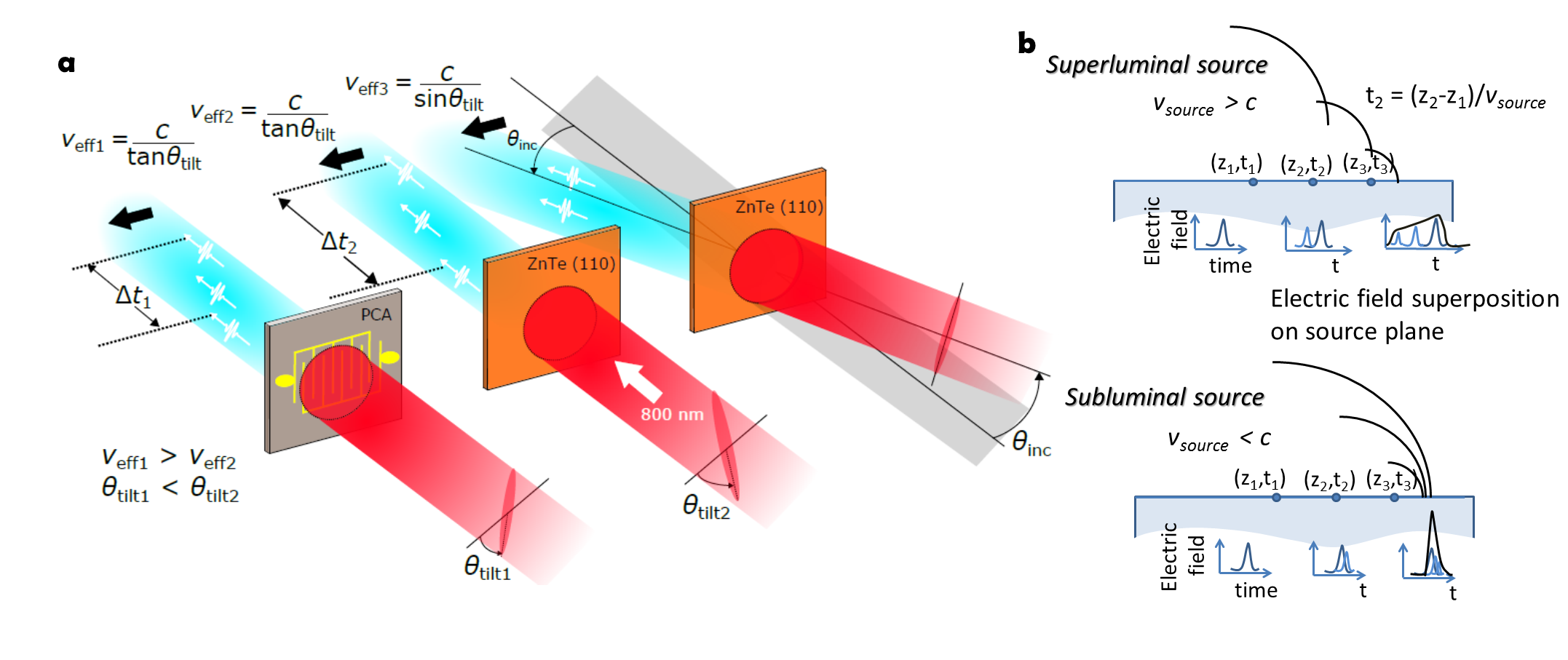}
\caption{{\bf (a)} Effective travelling wave concept, for both a non-linear  material and a photoconductive antenna excited in free space.
For a tilted pulse front the effective velocity is unbounded, while for an obliquely incident plane wave the effective source velocity is constrained to be greater than c. Tilt and incident angles, $\theta_{\rm tilt}$ and $\theta_{\rm inc}$ respectively, are those external to the source material. {\bf (b)} For a super-luminal source velocity the Huygens waves interfere to give a plane wave, while interference and cancellation is absent for a single-cycle 
sub-luminal source.  }
\label{TravelingWaveConcept:Fig}
\end{figure*}

To obtain acceleration over extended interaction lengths requires generation of electromagnetic modes with longitudinal electric field polarization, and real or effective phase matching of the electromagnetic carrier wave with the $\beta<1$ velocity of the relativistic particles. Longitudinal polarized terahertz radiation can be obtained through coherent spatial addition of phase or polarity offset beams \cite{Cliffe2016}, through selective coupling into (or onto) geometrically defined structures \cite{Deibel2006,Grosjean2008}, using air-plasma generation \cite{Minami2013}, using a segmented non-linear generation crystal \cite{Imai2012}, or using radially biased photoconductive antennas \cite{Cliffe2014,Kan2013,Waselikowski2013,Winnerl2012,Winnerl2009}. The phase-matching problem is however more fundamental; free space beams such as TEM$_{10}$ Gaussian modes necessarily have phase velocities greater than the vacuum speed of light, $c$, with the Guoy $\pi$-phase shift ultimately limiting available interaction lengths to less than a Rayleigh length\,\cite{Esarey1995}. Waveguiding is capable of obtaining $\beta<1$ phase matching at a single frequency (or subset of frequencies) but such structures are necessarily subject to dispersion relations that preclude maintaining the single-cycle field profile over an extended region.  For few-cycle ($\lesssim 10$\,ps) THz pulses, phase matching and high field strengths can be maintained over longer distances, but they are subject to a further inherent limitation of separation of electromagnetic energy and particle location through group-phase velocity mismatch \cite{Wong2013}.

Here we demonstrate the generation of terahertz-frequency electromagnetic fields that are single-cycle, polarized in the direction of propagation, that propagate in vacuum without dispersion, and have phase velocities that can be tuned to less than the velocity of light. The concept of velocity and polarization control is demonstrated with differing physical mechanisms of THz pulse generation, photoconductive antenna and non-linear optical rectification, and in non-linear media known to be capable of fields strengths exceeding the 10\,MV/m of start-of-the-art superconducting accelerators.
 The concept is scalable without limit in interaction length, not being subject to limitations of dispersion, and we show that it is possible to achieve sustained energy gain over an extended distance. It is shown that the transverse forces accompanying the acceleration fields provide guiding focusing fields analogous to those in RF cavities. Our approach will enable the length scalable high-gradient relativistic particle acceleration required for application to femtosecond electron diffraction, and for driving compact x-ray free-electron lasers.
\\
\\
{\bf\normalsize Concept description} \\
{\bf Sub-luminal travelling wave.}
Laser-driven THz sources based on photoconductive antenna (PCA) or optical rectification provide a conversion of the optical pulse (group) arrival time to a THz carrier phase.
For the PCA the optical energy deposition is responsible for carrier generation, setting the time (phase) of the radiating surface current surge~\cite{Jepsen1996}. For optical rectification, the coherent convolution of difference frequency mixing combinations can be shown to produce a THz carrier field following the time-derivative of the optical envelope \cite{Ahn2003}.
To achieve a sub-luminal carrier wave we exploit this group-to-phase conversion together with tilted optical pulse-fronts giving a controllable arrival-time delay exciting a planar THz source. Tilted pulse-fronts are generated through imaging of the pulse propagated from an optical diffraction grating. The optical wavefronts of the pulse remain orthogonal to the propagation direction, while there is a transverse time delay in the local energy content of the pulse.
As shown in Fig.~\ref{TravelingWaveConcept:Fig} when impinging on a source plane (a non-linear $\chitwo$ material or PCA) the delayed optical energy arrival produces an effective travelling wave source for the THz. 
In the case of optical rectification, the preservation of wave-front direction orthogonal to propagation allows the $\chitwo$  phase-matching conditions to be maintained independent of the pulse-front-tilt angle.
The use of tilted pulse-fronts is central to obtaining an effective sub-luminal THz phase velocity. With pulse-front tilt spanning the range $0\leqslant \phi_{\rm tilt} <\pi/2$ the effective surface optical group velocity (and hence THz phase velocity) can span $0 < v^{\rm opt, eff}_{\rm group} \leqslant \infty$. As the optical carrier wavefront remains travelling with normal angle of incidence with effective surface velocity $v_{\phi}^{\rm opt, eff}=\infty$ it cannot however be exploited directly for acceleration. 
While a simpler non-normal incident planar pulse front can also provide effective velocity control, such an arrangement is limited to velocities greater than $c$; due to Snell's law refraction of the pulse front on entry to the velocity reducing medium, this limitation of planar pulse-fronts holds even when the source is embedded in a material of refractive index $n>1$.

{\bf Propagating `evanescent' waves.} 
The excitation of the THz sources through tilted optical pulse fronts allows coupling of the optical energy into the source material at sub-luminal velocities. The outcoupling of the THz pulse from inside the source material is likewise subject to conditions of boundary continuity and refraction, and for conventional many-cycle electromagnetic waves the transition to sub-luminal source propagation is equivalent to meeting conditions for the critical-angle of total internal reflection and post-boundary evanescent wave propagation \cite{Mochan2001}.
For the single cycle pulses generated by optical rectification the classification of the post-boundary fields as non-propagating and exponentially decaying in amplitude is no longer appropriate. Conceptually describing the travelling-wave source as a superposition of time delayed Huygens spherical waves the  sub-luminal and super-luminal velocities, and single-cycle versus multi-cycle pulses, give rise to qualitatively different solutions for the propagating waves. For the super-luminal source (Fig.~\ref{TravelingWaveConcept:Fig}b) a superposition of waves gives rise to a solution with a far-field distribution approximating an obliquely propagating plane-wave satisfying Snell's law of refraction. For the sub-luminal case the Huygens waves of a single-cycle pulse cannot coincide, with the exception being $v_{\phi}=c$ for which they add coherently along the surface. For single-cycle pulses the process of constructive or destructive interference for $\Delta \phi >2\pi$ does not arise and the exponential decay of the field with distance from the surface associated with evanescent waves is supplanted with a slower $1/r$ decay of the Huygens fields and a temporal stretching arising from the superposition of wavelets retarded in time by the sub-luminal source velocity. 


\begin{figure}
\includegraphics[width=9cm]{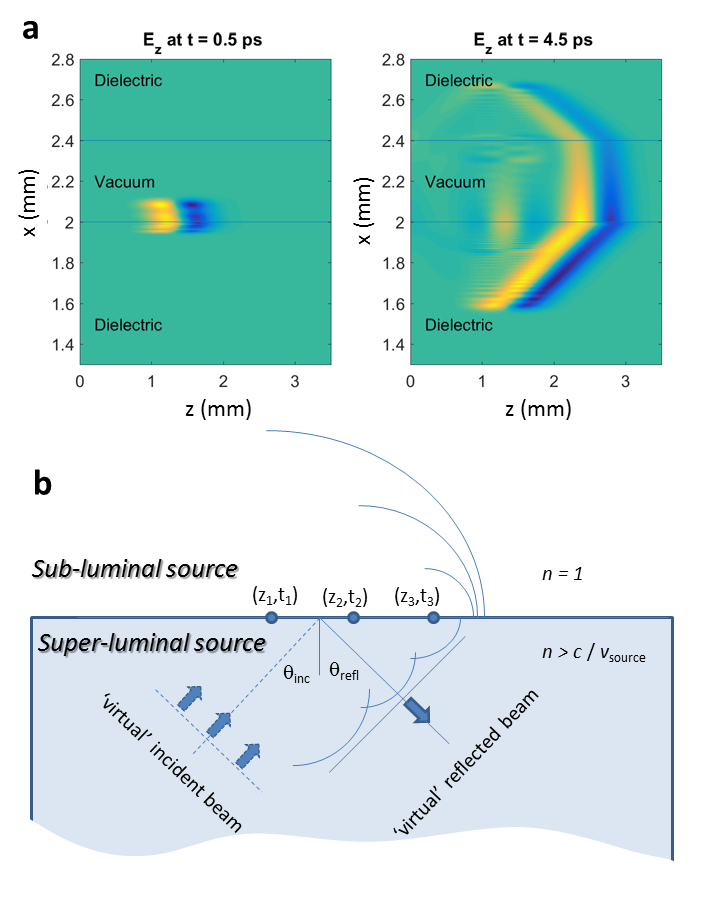}
\caption{{\bf (a)}. Finite difference time domain calculations of the electric field for a travelling source on the boundary of a dielectric. The source velocity is grater than the speed of light within the dielectric, but less than the speed of light in vacuum. {\bf (b)} The travelling source can be viewed as a surface excitation driven by a virtual incident plane wave. For a sub-luminal (in vacuum) effective surface velocity the angle of incidence for the virtual wave will exceed the critical angle for total internal reflection. }
\label{TotIntReflection:Fig}
\end{figure}

To provide a detailed and quantitative picture of the single-cycle propagation from a sub-luminal source, finite difference time domain (FDTD) simulations have been undertaken (Methods: FDTD simulations). An example of finite difference time domain calculation of emission from (and into) the surface under sub-luminal conditions is presented in Fig.~\ref{TotIntReflection:Fig}. After an initial stage of propagation where the field is established in the region above the source plane, a stable pulse is obtained, travelling with a wavefront normal to the surface and with a velocity set by the effective source velocity. The normal-to-surface extention of the field structure, which evolves over a finite time, is responsible for an apparent instananeous propagagtion of the field across a gap in frustrated total internal reflection \cite{Mochan2001,Carey2000}.
The physical connection to total internal reflection is described schematically in Fig.~\ref{TotIntReflection:Fig}b, and is apparent in the numerically calculated field profiles of Fig.~\ref{TotIntReflection:Fig}a. A surface sub-luminal travelling source is analogous to that arising from a virtual  planar wavefront arriving at an angle of incidence exceeding the critical angle. This virtual incident beam establishes a real `reflected' beam together with the sub-luminal fields propagating across the boundary into the vacuum region. 
The use of total internal refection and `evanescent' fields have been previously proposed for particle acceleration~\cite{Frandsen2006,Palfalvi2014}.
While the resulting field structures and evolution is analogous to that expected in total internal reflection, our  scheme does not rely on a real (THz) incident beam and avoids the associated obstacles of in-coupling of a pulse at a sub-critical angle.
\\
\begin{figure*}[htb]
\includegraphics[width=16cm]{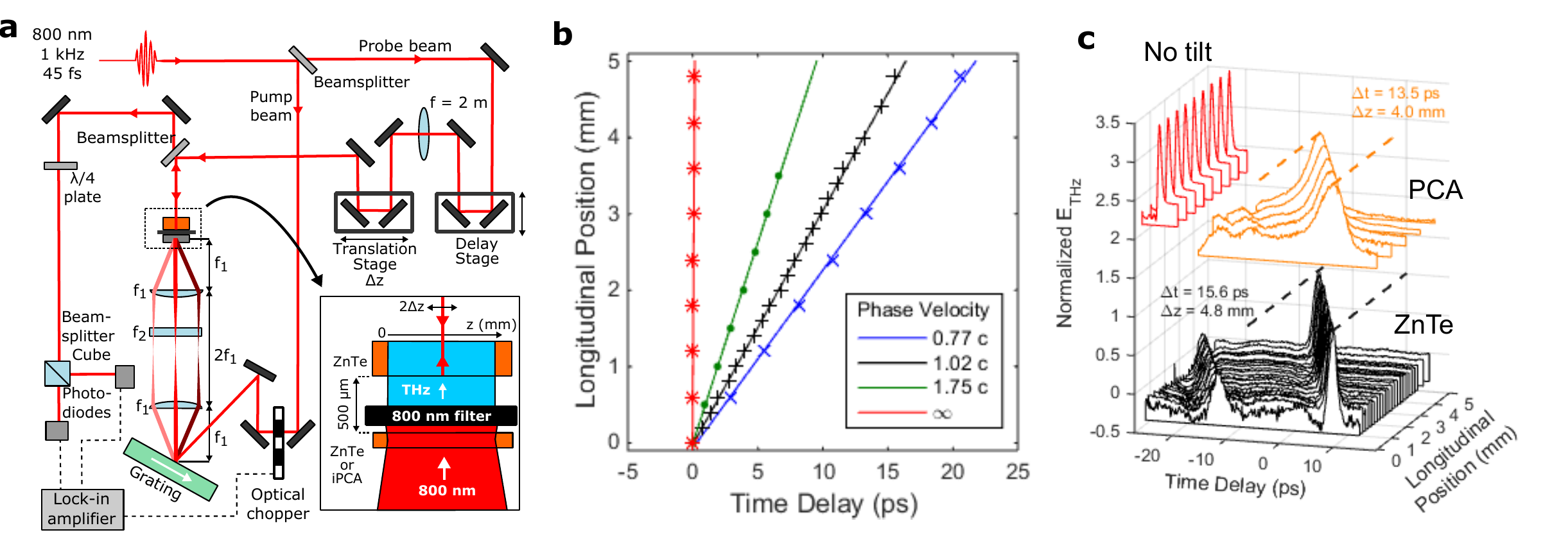}
\caption{(\textbf{a}) Experimental arrangement for the spatial and temporal characterisation of the THz pulse emitted from a ZnTe or iPCA travelling wave source.
(\textbf{b}) World lines of the THz pulse measured for a ZnTe source with differing pulse-front tilts.
(\textbf{c}) Examples of the single-cycle THz pulse pulse temporal profiles measured at positions in a plane parallel to the emitting surface. The longitudinal direction is defined as the direction of propagation of the travelling wave source. From top to bottom, the pulses were generated by a planar wavefront optical pulse exciting a ZnTe emitter ($v^{\rm eff} = \infty$), and $\theta_{\rm tilt} \sim 45^{o}$ pulse front exciting an iPCA and ZnTe emitters.}

\label{dispesionlessExamples:Fig}
\end{figure*}
\\
{\bf\normalsize Results} \\
{\bf Observation of single-cycle sub-luminal propagation.} Sub-luminal dispersionless THz pulses have been generated using the concepts of Fig.~\ref{TravelingWaveConcept:Fig}, in both a large-area interdigitated photoconductive antenna and through optical rectification in a (110)-cut ZnTe single crystal (see Methods: THz Generation). The experimental configuration is shown schematically in Fig.~\ref{dispesionlessExamples:Fig}a. In both cases the optical pulse (group) front tilts were produced by a diffraction grating, with the diffracted pulse imaged onto the generation PCA or ZnTe crystal. The optical pump beam was incident normal to the generation plane. Temporal and spatial characterization of the field was undertaken with electro-optic sampling with a 50\,fs optical probe retro-reflected from the internal boundary of a separate ZnTe electro-optic detection crystal. The THz electric field emitted from  the source and propagated through air into the detection material effectively creates a birefringence in the detection crystal that temporally and spatial replicates the incident THz pulse, and this birefringence is observed through the polarization change of the optical probe (see Methods: EO Detection).
Fig.~\ref{dispesionlessExamples:Fig}c shows pulse propagation measurements for a normal-incidence planar pulse front  optical pump on a ZnTe emitter, with $v_{\rm eff}^{\rm THz} = \infty$, and measurements of $v_{\rm eff}^{\rm THz} \approx c$ pulse propagation in ZnTe and PCA sources. The tuneability from sub- to super-luminal propagation has been observed with temporal-spatial mapping of the pulse evolution undertaken for a range of pulse-front tilts, corresponding to effective velocities from $1.75c$ down to $0.77c$. Measured world-lines in ZnTe are shown in Fig.~\ref{dispesionlessExamples:Fig}b.  We observe no significant pulse broadening or reshaping as a function of wave-propagation along the surface; variations in intensity are attributable principally to the transverse intensity profile of the optical pulse providing the THz excitation.

\begin{figure*}
\includegraphics[width=18cm]{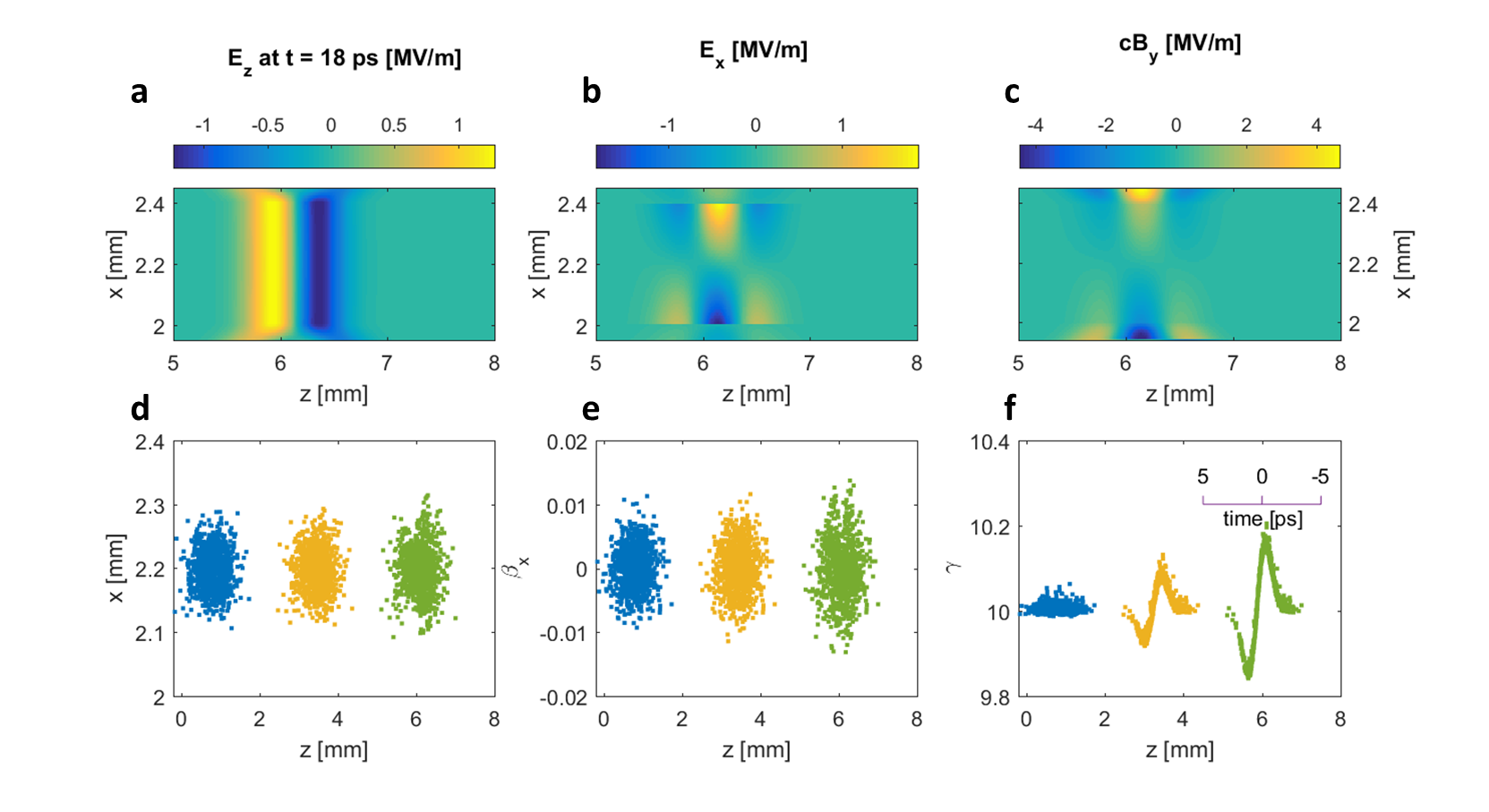}
\caption{Particle acceleration in a paired travelling-source structure. (\textbf{a-c}): longitudinal electric field ($E_{z}$) and transverse electric and magnetic field components ($E_{x}$, $H_{y}$) for a 500\,fs, 10\,MV/m THz source field generated in one crystal. The vacuum/dielectric interfaces are marked with dotted lines.  (\textbf{d-f}) Transverse position, angle and  energy respectively for 1000 electrons after acceleration in a 15\,mm structure as a function of particle position in the bunch; particles enter the structure at 5.2\,MeV/c, with 500\,fs rms duration with 100\,$\mu$m rms diameter. }
\label{DoubleSide:Fig}
\end{figure*}

{\bf Application to particle acceleration and manipulation.} 
Both PCAs and appropriately orientated $\chitwo$ optical rectification sources will produce THz electric fields polarized in the plane of the source. With the field polarization aligned in the  direction of propagation of the travelling wave (the longitudinal direction) co-propagating charged particle acceleration becomes possible.
The longitudinal accelerating electric fields  are accompanied by significant transverse electric and magnetic fields which arise through the spatial and temporal gradients of the THz source. For a relativistic particle with $\beta \lesssim c$ the magnetic deflecting force becomes comparable to the accelerating force. Additionally the transverse electric field necessary to satisfy the condition of vanishing electric field divergence
 in the vacuum gap becomes comparable to the accelerating field temporally ahead and behind the region of acceleration. As in laser-dielectric acceleration schemes, and in conventional RF cavity acceleration, these deflection forces can be eliminated or reduced through imposing additional symmetry around the particle beam axis. Here we consider a two dimension arrangement with sources located in a pair of $z-y$ planes. This two dimensional symmetric arrangement and the resultant field structure is similar to the total-internal reflection scheme
 proposed by Frandsen at al.\,\cite{Frandsen2006} and P\'{a}lfalvi et al.\,\cite{Palfalvi2014}. Figure~\ref{DoubleSide:Fig} shows the electric and magnetic fields, calculated with FDTD modelling, in the vacuum gap between a pair of opposing sources embedded in dielectric media separated by $400~\mu$m (see Methods: FDTD simulations). The longitudinal electric field sources $E_z(x=\pm250\,\mu{\rm m}, z-\beta_s t,t)$  are localized in $z-y$ planes with $x=\pm250\,\mu{\rm m}$, and propagate with source velocity $c\beta_s$. 
  The source velocity in these calculations was 
 $\beta_s = 1-5\times 10^{-3}$, correspond to velocity matching for a 5\,MeV electron beam. 
Due to cancellation of the transverse electric and magnetic fields in the symmetry plane of the source-pair, the deflection forces are minimal in a region of $\approx 100\,\mu$m around the central beam-axis, providing an acceleration potential bucket sufficient for injection of particle beams from conventional electron guns. 
 The field structure produced from the opposing pair travelling sources is similar to that found within conventional RF accelerating structures, with the exception that the accelerating fields and electromagnetic energy are co-propagating in synchronism with the charged particles rather than stored over the longer times and spatial extent in an accelerating cavity structure.
  
The results of modelling particle acceleration of a 5\,MeV electron beam injected into the travelling source paired structure are shown in Fig. \ref{DoubleSide:Fig}d-f.  The particle dynamics of have been determined through numerical solution of the relativistic Lorentz force equation, and include the electric and magnetic deflection forces (see Methods: FDTD simulations). For the two dimensional arrangement of the source, the acceleration and deflection occur in the 
 $z$ and $x$ axes, with $E_y, H_x, H_z = 0$ by symmetry. 
The injected beam has a $\sigma_t=1$\,ps rms pulse duration, and a transverse phase-space emittance of 0.3\,mm.mrad, consistent with that obtainable from state-of-the-art RF photo-injectors~\cite{EmmaLCLS2010,Vinatier2015}.
We show the evolution of the particle distribution as the particles interact over approximately 7\,mm with a  peak accelerating gradient of 10\,MV/m and a 500\,fs bi-polar THz field representative of our LiNbO$_3$ source to be described in the following section.  Nett energy gain of approximately 100\,keV is obtained when injecting close to the peak phase of the accelerating field.  
In contrast to acceleration schemes using optical wavelengths, the approximately 100\,$\mu$m longitudinal dimension of the acceleration bucket  ensures negligible phase slippage of  electron bunches with a typical $ 10^{-3}$ energy spread, while the 100\,$\mu$m transverse dimensions of the accelerating field allow for the entire injected electron bunch to be captured and accelerated without traverse phase space degradation.

\begin{figure*}
\includegraphics[width=17cm]{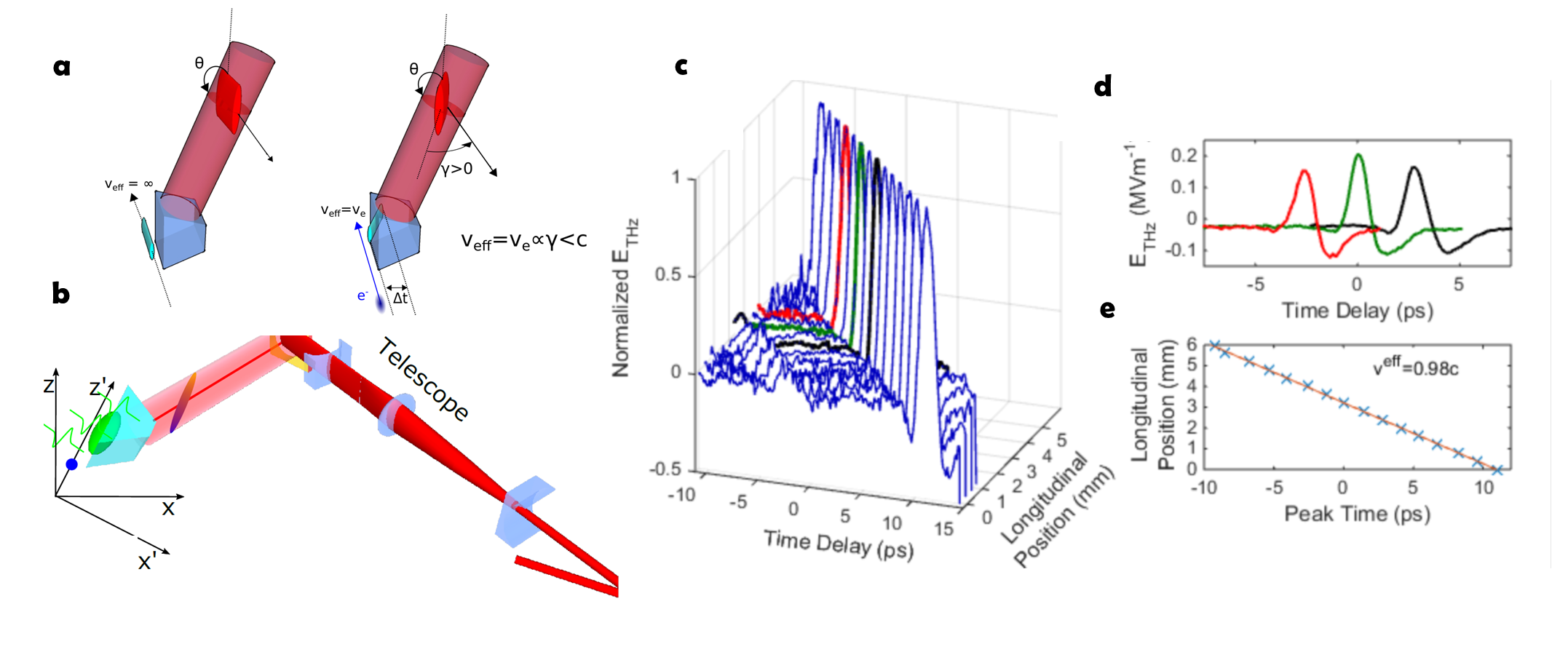}
\caption{Pulse-front double-tilt concept for high-field strength travelling source THz generation in \LN. \textbf{(a)}  arrangement for generating a pulse tilt satisfying the Cherenkov phase matching, and simultaneously with a travelling source co-propagating with an electron bunch. The longitudinal direction is defined as the direction of propagation of the travelling wave source and electron beam. \textbf{(b)}  for the \LN\  measurements the pulse-front is titled in the laboratory $x-y$ frame, and with the \LN\ crystal orientated to project the tilt into orthogonal Cherenkov and travelling-source directions. 
\textbf{(c-e)} Measured THz pulse propagation in a plane parallel to the \LN\ emitting surface.}
\label{LNdata:Fig} 
\end{figure*}

{\bf MV/m field strengths for particle acceleration.}
 The above measurements with interdigitated PCA and ZnTe optical rectification sources serve to illustrate the concept of a single-cycle source with longitudinal polarization and sub-luminal dispersionless propagation. Neither of these optical to THz mechanisms or media are however capable of generating the multi-MV/m field strengths sought for high-gradient relativistic particle acceleration. To achieve high field strengths the same travelling source concept has been adapted to work with a LiNbO$_3$ non-linear medium, a crystal that is capable of  greater than 100\,MV/m field strengths\,\cite{Hirori2011}.
A significant hurdle is encountered in using this established high-field THz material, in that non-colinear THz and optical propagation is required to satisfy the THz and optical phase-matching\,\cite{Fulop2011}.
For LiNbO$_3$ the optical group velocity exceeds the phase velocity of the generated THz radiation by a factor of approximately 2.5. As a consequence the THz radiation is generated in an obliquely propagating  Cherenkov cone centred around the laser propagation axis. To allow efficient generation from a transversely extended optical beam a temporal-spatial correlation is introduced in the optical excitation pulse, with the correlation such that the locally produced Cherenkov cones add constructively in a specific Cherenkov direction\,\cite{Fulop2011}. 
To additionally provide a  $v_{\phi}^{\rm eff}<c$ travelling source, we have developed a `double-tilt' pulse front scheme as shown in Fig.~\ref{LNdata:Fig}. 
In this scheme a single diffraction grating generates a tilt exceeding that required for the transverse Cherenkov  phase matching; the non-linear LiNbO$_3$ medium is then orientated such that when geometrically 
projected onto the THz emission surface the tilt is resolved into a Cherenkov-matching tilt and an orthogonal travelling-source tilt. 
The optical phase fronts remain orthogonal to the optical direction of propagation, and optical waveplates provide polarization adjustment so that
the refractive index ellipsoid and 3-wave phase matching conditions, and the 
 $\chitwo_{ijk}$ tensor orientation and emitted THz polarization, remain unchanged from the commonly employed LiNbO$_3$ THz generation configuration. 

The double-tilt approach has been demonstrated through an optical arrangement similar to that employed for the ZnTe and PCA measurements. The combined pulse-front tilt was maintained in the horizontal lab-frame horizontal plane, with the consequence that the transverse and longitudinal (effective propagation) directions, and the emission surface of the \LN\ are oriented out of the lab-frame horizontal or vertical  planes. As for the ZnTe and PCA measurements, a temporal and spatial mapping of the fields near the surface was obtained with EO sampling.
The transverse intensity profile of the optical beam is also shaped with cylindrical telescopes so that following oblique incidence on the grating and  projection onto the out-of-plane oriented crystal surface, the THz source dimensions are approximately 15\,mm in the propagating source longitudinal direction, and 2\,mm in the orthogonal transverse (Cherenkov matched) direction. 

A temporal and spatial map of the THz field 500\,$\mu$m from the \LN\ surface was measured with electro-optic sampling and is summarised in the data of Fig.~\ref{LNdata:Fig}c-e. In this data set the single-cycle THz pulse is  travelling at 0.98c, as determined through the time-delay of the peak longitudinal electric field as a function of longitudinal position across the emitting surface.
The laser power incident on the \LN\ THz generating crystal was limited by the available laser system, and diffraction grating and optical losses, to 600\,$\mu$J. The transverse optical pulse shaping increased the optical pump energy density in the non-linear \LN\ while maintaining the desired propagation distance of approximately 10\,mm. 
Absolute electric field strength of a THz pulse can in principal be determined in EO sampling, through calibration of the EO signal into absolute polarization rotation of the optical probe together with knowledge of the phase-matching and $\chi^{(2)}$ non-linear response of the EO detection crystal (see, for example, Cliffe et al.\cite{Cliffe2014}). Where the THz and optical probe do not propagate collinearly in the detection crystal, the
experimental response 
 function is further subject to angular separation of the optical probe and non-linear generated optical 
 waves~\cite{Walsh2014}. 
 The physical separation of the optical waves in propagation from detection crystal to the photo-diode detectors will give rise to an underestimate of the THz electric field strengths. The field reduction arising from this physical separation of optical waves is extremely challenging to objectively determine, and here we have instead provided a lower-bound measurement of the electric field, based on an assumption of collinear THz and optical probe waves. With our modest laser powers we obtain a lower bound measured field strength of 0.20\,MV.m$^{-1}$. Based on established capability of \LN\ we believe the true field strength to significantly higher, and that with improvements in laser system field strengths approaching 10\,MV.m$^{-1}$ may be readily obtained.

{\bf\normalsize Discussion}\\
We have described and demonstrated  a method for generating single cycle THz pulses that propagate with an effective sub-luminal velocity, and without distortion of the single-cycle field profile during propagation. The electric field polarization is constructed with a component in the longitudinal or propagation direction as required for acceleration of charged-particle beams. The transverse electric and magnetic fields that arise from  the spatial and temporal gradients of the travelling source provide a focusing structure analogous to RF particle accelerators, but with a 1000-fold reduction in spatial scale. The 1\,ps duration of the THz pulse is 6-orders of magnitude smaller than the $\mu$s cavity fill-time of conventional RF accelerating structures. The $10^6$ reduction in the time-integrated energy content of the electro-magnetic field while maintaining field overlap with particle beams offers the potential for dramatic reductions in the power requirement of high energy particle accelerators.

\small

\section{Methods}

{\bf Pump Source.}
The Ti:Sapphire regenerative amplifier laser system used to generate THz radiation produced pulses with an energy of up to 2 mJ and a pulse duration of approximately 45\,fs at a repetition rate of 1 kHz, with a central wavelength of 800 nm. Three different generation media were used; a 0.5 mm thick (110)-cut ZnTe crystal, a commercially-available interdigitated photoconductive antenna (TeraSED10, Laser Quantum Ltd.), and a 6\% magnesium-oxide doped stoichiometric lithium niobate crystal. Further details of the interdigitated photoconductive antenna (iPCA) can be found in Ref.~\cite{Beck2010}. 

{\bf THz Generation: iPCA and ZnTe.}
The interdigitated photoconductive antenna was biased with 25 V pulses at a repetition rate of 500 Hz with a duty cycle of approximately 10\% and photo-excited with 25\,$\mu$J pulses at a repetition rate of 1 kHz from the regenerative amplifier laser system.  The ZnTe crystal was pumped with 500\,$\mu$J pulses and an optical chopper was used to reduce the repetition rate to 500\,Hz for lock-in detection.  A filter was used to block the 800\,nm pump light from being transmitted to the ZnTe detection crystal. Diffraction gratings with 1200\,lines/mm and 600\,lines/mm were used in a 4f imaging arrangement (consisting of two 250\,mm focal length cylindrical lenses) to produce a range of pulse-front-tilt angles within the generation medium.  A 400\,mm cylindrical lens was used to achieve an increase in the excitation energy density at the terahertz generation point without affecting the pulse-front-tilt angle.

{\bf THz Generation: LiNbO$_3$.}
The lithium niobate crystal was pumped at 500\,Hz with 600\,$\mu$J pulses after losses from optics and the diffraction grating. A 1200 lines/mm diffraction grating was used to produce a pulse front tilt of 43.8$^o$ and was imaged to the crystal using a single lens giving a magnification factor of 0.21 to give a total pulse front tilt of 77.6$^o$ (63.8$^o$ inside the LiNbO$_3$). A zero-order half-wave plate was used to rotate the polarization of the first order diffracted beam in order to match the $z-$axis of the lithium niobate crystal. 
The crystal was rotated by an angle of approximately 12.7$^o$ to give a phase matching component of the tilt of 77.3$^o$ (63.0$^o$ inside the crystal) and a ‘velocity generating’ component of 45.0$^o$ (24.2$^o$ inside), giving an expected velocity of approximately 1.0 c. The pump beam was shaped into an ellipse with the major axis aligned along the direction of travel of the generated terahertz wave. This was achieved using a telescope consisting of a two -50 mm focal length cylindrical lenses and a 250\,mm spherical lens. These were used to produce a magnification factor of 5 in the vertical axis and 1/5 in the horizontal axis. The two cylindrical lenses were rotated about the pump propagation axis by approximately 3.6$^o$ in order to match the ellipse rotation to that of the lithium niobate crystal. The 3.6$^o$ rotation, along with projections onto the diffraction grating, produced a final elliptical beam with a rotation of approximately 12.7$^o$.

{\bf EO Detection.}
In the iPCA and ZnTe experiments, a 1 mm thick (110)-cut ZnTe crystal was used to detect the generated terahertz radiation with a separation of approximately 500\,$\mu$m between the generation medium and the detection crystal. In the LiNbO$_3$ experiments, a 0.5 mm thick (110)-cut GaP crystal was used with a separation of approximately 300\,$\mu$m. In both cases, a balanced electro-optical detection scheme was used to measure the generated terahertz electric fields. The reflection of the probe pulse off of the inner surface of the detection crystal acquired a polarization change which was proportional to the terahertz electric field strength. The probe pulse was then separated into two orthogonal polarization states using a polarizing beam splitter and the intensities of each were measured using balanced photodiode detection scheme. A lock-in amplifier was used to measure the voltage change from the two photodiodes at the 500 Hz frequency of the pump pulse or the iPCA bias voltage. The probe beam was focused to a $\lesssim 100\,\mu$m diameter spot at the detection crystal and its position could be translated horizontally and vertically, while maintaining a constant path length, using translation stages in order to measure the terahertz electric fields at different spatial positions over the crystal surface.
A calibration of the probe beam translation stages was performed by attenuating the probe pulse and taking images of the probe beam displacements on the surface of the detection crystal using a CCD camera.

{\bf FDTD simulations and particle evolution.}
The Finite Difference Time Domain (FDTD) method is used to determine the fields as a function of time. The travelling bipolar THz source is embedded within the dielectric material ($\epsilon = 9\epsilon_0$) with velocity$\beta_s = 1 - 5\times 10^{-3}$, matching a $\gamma=10$ (U $\approx$ 5\,MeV) electron beam velocity. In line with the optical rectification generation process there is no magnetic field source and the electric field of the source term is polarized purely in the $z-$direction, which also corresponds to our chosen source propagation direction. The spatial grid for the simulations of Figs.~\ref{TotIntReflection:Fig} and \ref{DoubleSide:Fig} is 5\,$\mu$m and the time step is 12\,fs

The particle phase space evolution was obtained through numerical solution of the relativistic equation of motion
\begin{equation}
\frac{{\rm d}\bm{\beta}}{ {\rm d}t} = -\frac{e}{m_{\rm e}c}\,\sqrt[]{1-\left|\beta\right|^{2}}\left[\mathbf{E}+c\bm{\beta}\times\mathbf{B}-\bm{\beta}\left[\bm{\beta}\cdot\mathbf{E}\right]\right]
\label{eq:lorentz}
\end{equation}
A fourth-order Runge-Kutta  algorithm was embedded within the FDTD field evaluation algorithm, with particle velocities and positions updated at each time step of the FDTD code. The  position and momentum of  1000 electrons  were tracked as they progated through the double-sided field structure.
 The initial `injected' electron beam parameters are chosen to be representative of the VELA linear electron accelerator \cite{VELA2013,VELA2015} with electron momentum $p$ = 5.2 MeV/c ($\beta$ = 0.996). 
 \\

{\bf Acknowledgements.}\\
This work was supported by the United Kingdom Science and Technology Facilities Council [Grant No. ST/G008248/1];  the  Engineering  and
Physical Sciences Research Council [Grant No. EP/J002518/1]; and the Accelerator Science and Technology Centre through Contract No. PR110140.
\\
\\
{\bf Author contributions.}\\
S.P.J. and D.A.W. proposed the concept and initial experiment, which was further developed by all. D.S.L. performed the experiments with help from D.A.W. and M.J.C., E.W.S. and S.P.J. performed the FDTD simulations. S.P.J. wrote the manuscript with contributions from all and revisions by D.M.G. and D.A.W., D.M.G. and S.P.J. supervised the project. 
\\
\\
{\bf Additional information}\\
Competing financial interests: The authors declare no competing financial interests.

The data associated with the paper are openly available from The University of Manchester eScholar Data Repository: http://dx.doi.org/xx.xxxxx/x.xxxxxx.

\label{lastpage}
\end{document}